\documentclass[
 reprint,
 amsmath,amssymb,
 aps,
 superscriptaddress,
]{revtex4-1}

\usepackage[section]{placeins} 
\usepackage{cancel} 
\usepackage{lineno} 
\usepackage[utf8]{inputenc} 
\usepackage{natbib} 
\usepackage{csquotes} 
\usepackage[caption = false]{subfig} 
\usepackage{graphicx}
\usepackage{dcolumn}
\usepackage{bm}
\usepackage{color} 
\DeclareMathOperator{\e}{e}
\graphicspath{{Figures/}} 
\usepackage{hyperref} 
\hypersetup{colorlinks=true, linkcolor=blue, filecolor=magenta, urlcolor=blue}  

\begin{document}

\title{Thermal weakening of cracks and brittle-ductile transition of matter: a phase model}

\author{Tom Vincent-Dospital}
\email{vincentdospitalt@unistra.fr}
\affiliation{Université de Strasbourg, CNRS, IPGS UMR 7516, F-67000 Strasbourg, France}
\affiliation{SFF Porelab, The Njord Centre, Department of physics, University of Oslo\\P. O. Box 1048, Blindern, N-0316 Oslo,  Norway}

\author{Renaud Toussaint}
\email{renaud.toussaint@unistra.fr}
\affiliation{Université de Strasbourg, CNRS, IPGS UMR 7516, F-67000 Strasbourg, France}
\affiliation{SFF Porelab, The Njord Centre, Department of physics, University of Oslo\\P. O. Box 1048, Blindern, N-0316 Oslo,  Norway}

\author{Alain Cochard}
\affiliation{Université de Strasbourg, CNRS, IPGS UMR 7516, F-67000 Strasbourg, France}

\author{Knut J\o rgen M\aa l\o y}
\affiliation{SFF Porelab, The Njord Centre, Department of physics, University of Oslo\\P. O. Box 1048, Blindern, N-0316 Oslo,  Norway}
\author{Eirik G. Flekk\o y}
\affiliation{SFF Porelab, The Njord Centre, Department of physics, University of Oslo\\P. O. Box 1048, Blindern, N-0316 Oslo,  Norway}

\date{\today} 
                              
\begin{abstract}
We present a model for the thermally activated propagation of cracks in elastic matrices. The propagation is considered as a subcritical phenomenon, the kinetics of which being described by an Arrhenius law. In this law, we take the thermal evolution of the crack front into account, assuming that a portion of the released mechanical energy is transformed into heat in a zone surrounding the tip.
We show that such a model leads to a two-phase crack propagation: a first phase at low velocity in which the temperature elevation is of little effect and the propagation is mainly governed by the mechanical load and by the toughness of the medium, and a second phase in which the crack is thermally weakened and propagates at greater velocity. Such a dual behavior can potentially explain the usual stick-slip in brittle fracturing, and we illustrate how with numerical simulations of mode I cracks propagating in thin disordered media. In addition, we predict the existence of a limiting ambient temperature above which the weakened phase ceases to exist and we propose this critical phenomenon as a novel explanation for the brittle-ductile transition of solids.
\end{abstract}     
                         
\maketitle

\section{Introduction}

Of paramount importance in engineering and geophysics, the impact of temperature in fracturing processes have since long been studied. It can simplistically be sorted into two categories: background effects where the temperature is treated as an environmental constant affecting the rates at which the defects of a medium are propagating or healing \cite{Brenner, Zhurkov1984, creepBaud, lawn_1993} and dynamic effects where the propagation of fractures self-induces a rise in temperature in the vicinity of the crack front \cite{RiceLevy,Fuller1975,tribo_blackbody,Bouchaud2012,ToussaintSoft}. In the latter case, the heat elevation can be regarded as more than a secondary effect of the medium's damage: it can be an active process back affecting the crack propagation. This phenomenon will be here referred to as \textquote{thermal weakening.} Such a weakening has notably been studied in earth science where it is believed to play a role in faults stability and earthquake triggering \cite{ThermalRunaway, HeatWeak} and it was included in the so-called rate-and-state framework \cite{ratestate} as an explanation for rate weakening faults. Several mechanisms have been proposed to explain thermal weakening, such as the softening \cite{Marshall_1974, carbonePersson} or melting of fracture surfaces or the thermo-pressurization of fault fluids \cite{FluidPresMelt,pressur2005,SulemCarbo}. We here consider a model which disregards such effects and focuses on the statistical physics consideration of higher reactions rates (i.e., quicker fracture propagation) at higher temperatures, as implied by an Arrhenius law \cite{kinetics}. This model notably showed good agreement with the rupture dynamics, experimentally reported in various polymers \cite{TVD2}. In this work, we further discuss how, in addition, it stands as a physical explanation for the brittle-ductile transition of matter.

\section{The thermal weakening model}

\begin{figure}
  \includegraphics[width=1\linewidth]{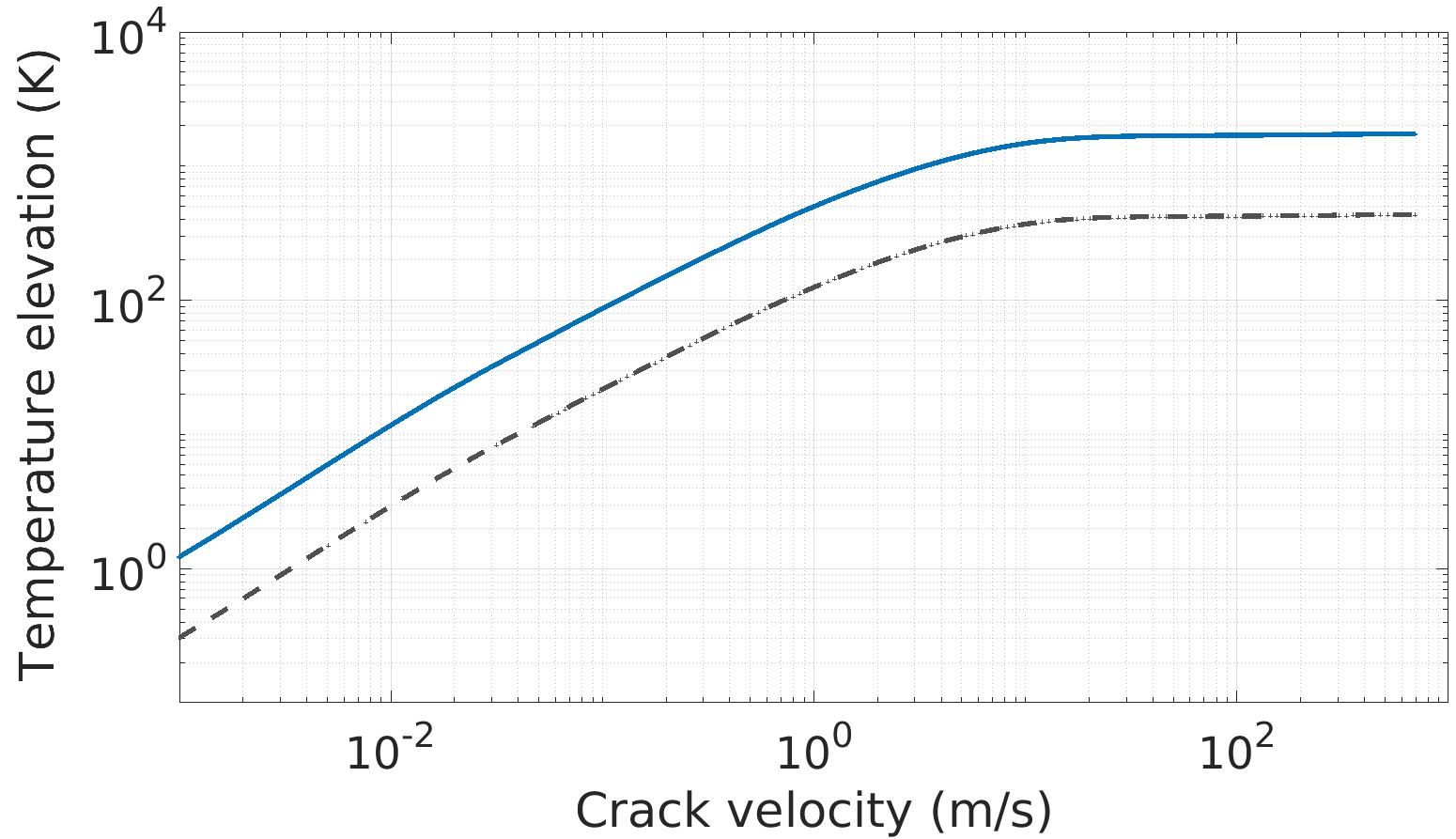}
  \caption{Steady state values of the temperature elevation. It is obtained by solving Eq. (\ref{tempdiff}) for a crack propagating at constant velocity and for $\phi G = 200$\,J\,m\textsuperscript{-2} (plain plot) and $\phi G = 50$\,J\,m\textsuperscript{-2} (dotted plot).}
  \label{ttovel}
\end{figure}
Arrhenius based models for the velocity of crack fronts have long been considered \cite{Brenner, Zhurkov1984, Santucci2004, lawn_1993} and have recently been shown to show good agreement with experimental observables of mode I cracks slowly propagating in acrylic glass bodies \cite{Lengline2011,Tallakstad2011,Cochard}. The rupture is then not considered as a Griffith-like threshold mechanism \cite{Griffith1921} where the crack only advances for $G>G_c$, where $G$ is the energy release rate of the crack in J m\textsuperscript{-2} (arising from the mechanical load given to the crack front) and $G_c$ the fracture energy of the medium (the energy barrier per surface unit to overcome molecular bonds). It is rather considered as a thermally activated subcritical phenomenon ($G<G_c$) for which the crack velocity is expressed as:
\begin{equation}
   V = \alpha \nu \e^{\cfrac{\alpha^2(G-G_c)}{k_B T}}
   \label{arrh1}
\end{equation}
where $\alpha$ is a characteristic size (m) of the fracturing process, that is associated with its energy barrier. $k_B \approx 1.38\times {10}^{-23}$\,J\,K\textsuperscript{-1} is the Boltzmann constant, $T$ the absolute temperature at the crack tip and $\nu$ the thermal bath collisional frequency.
Equation (\ref{arrh1}), as any Arrhenius law, is a continuous expression of a discrete process arising at the molecular scale. \citet{Cochard} have recently discussed it at length. The exponential term is the probability (i.e. $<1$) for the thermal agitation to exceed the activation energy $-\alpha^2(G-G_c)$ and hence for the crack to advance by a length $\alpha$. This probability is challenged every $1/\nu$ seconds. In theory $\nu$ is also temperature dependent but this is of negligible effect compared to the exponential dependence of the probability term \cite{kinetics} and we hence define $V_0=\alpha \nu$, the maximum crack velocity obtained when the activation energy is always reached. $V_0$ shall typically be in the range of the Rayleigh surface wave velocity \cite{Freund1972}. Because we consider the thermal evolution around the crack tip we also note $T=T_0+\Delta T$, where $T_0$ is the ambient temperature and $\Delta T$ any variation away from it at the tip.\\
Such variations are induced by the dissipation of the mechanical energy given to the elastic matrix in a plastic zone that surrounds the crack tip \cite{Irwin1957}. There are many processes responsible for such an energy loss, as the creation of new defects surfaces and the emission of mechanical waves, but we here focus on the release of heat. The model we use is based on the work of \citet{ToussaintSoft}: a portion $\phi$ of the energy release rate is dissipated on a cylindrical zone of radius $l$ centered around the crack tip. Such a configuration leads to a thermal evolution governed by:
\begin{equation}
   \frac{\partial (\Delta T)}{\partial t} = \frac{\lambda}{C} \nabla^2 (\Delta T) + \frac{\phi G V}{C \pi l^2}f
   \label{tempdiff}
\end{equation}
which is a diffusion equation including a source term. $\lambda$ is the medium's thermal conductivity in J\,s\textsuperscript{-1}\,m\textsuperscript{-1},K\textsuperscript{-1}, $C$ is the volumetric heat capacity in J\,K\textsuperscript{-1}\,m\textsuperscript{-3}, $t$ is the time variable and $\nabla^2$ is the Laplace operator. $f$ is the support function of the heat production zone of surface integral $\pi l^2$ (i.e., $f=1$ in the zone and $f=0$ otherwise). Solving this equation for a crack propagating at a constant velocity and constant release rate, one can show that the thermal elevation at the tip reaches a steady state after a short transient time. Figure \ref{ttovel} shows the evolution of this steady state as a function of $V$ and for two values of $G$. See the supplemental material for details on its computation. In our model, we use this relation to describe $\Delta T(V,G)$, thus discarding any transient regime. Equation (\ref{arrh1}) becomes:
\begin{equation}
   V = V_0 \e^{\cfrac{\alpha^2(G-G_c)}{k_B[T_0+\Delta T(V,G)]}}.
   \label{arrh3}
\end{equation}

\subsection*{Parameters used for illustration}

Note that most of the previously introduced parameters are strongly dependent on the medium in which the crack propagates. The figures we display here use parameters that could be likely for the propagation of interfacial cracks in sintered acrylic glass bodies \cite{Lengline2011,Tallakstad2011} and are discussed in the supplemental material: $\alpha=2.5\times {10}^{-11}$\,m, $G_c=250$\,J\,m\textsuperscript{-2}, $T_0=293$\,K, $C=1.7\times {10}^{6}$\,J\,K\textsuperscript{-1}\,m\textsuperscript{-3}, $\lambda=0.19$\,J\,s\textsuperscript{-1}\,m\textsuperscript{-1}\,K\textsuperscript{-1}, $V_0=1000$\,m\,s\textsuperscript{-1}, $l=20$\,nm and $\phi=1$. Note that we use this set of values only to propose some likely orders of magnitude for our parameters, and not to accurately represent the rupture of a specific material, as done in \cite{TVD2}.

\section{Phase behavior}

\begin{figure}
  \includegraphics[width=1\linewidth]{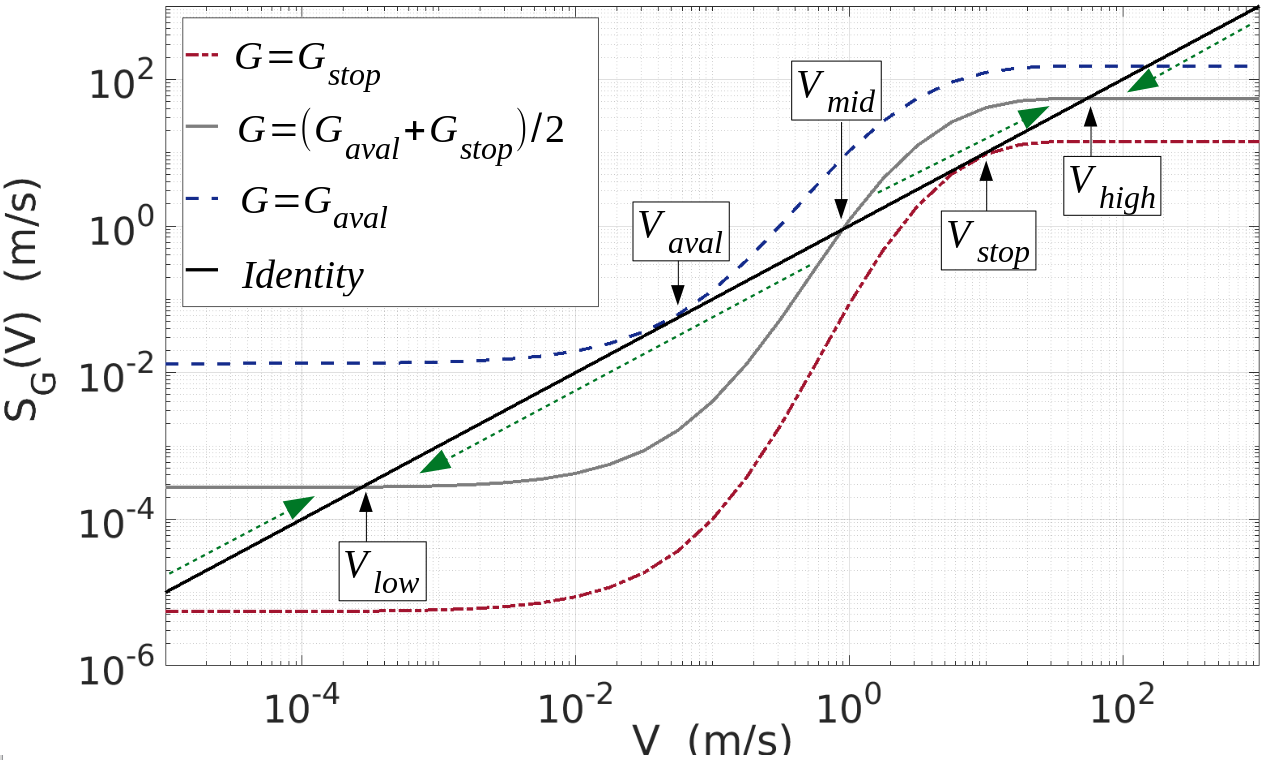}
  \caption{Representation of $V=S_G(V)$ for three values of $G$: $G_\text{stop}$, $G_\text{aval}$ ($>G_\text{stop}$) and the mid-value between $G_\text{stop}$ and $G_\text{aval}$. The intersections of $S_G$ with the identity plot (straight line) give the possible crack velocities. They are denoted $V_\text{low}$, $V_\text{mid}$ and $V_\text{high}$ and are emphasized for the intermediate $G$-plot. $V_\text{aval}$ and $V_\text{stop}$ are indicated on the two others plots. The dashed arrows indicate how off-balanced situations evolve to a stable fixed point.}
  \label{VSV}
\end{figure}
\begin{figure}
\includegraphics[width=1\linewidth]{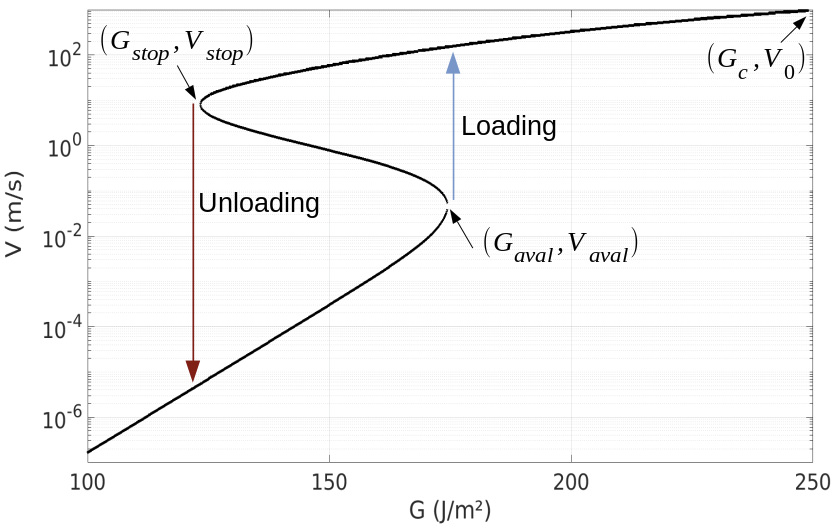}
  \caption{Solutions for the crack velocity as a function of $G$ for $T_0=293$\,K. All solutions in between $V_\text{stop}$ and $V_\text{aval}$ are unstable, any other point is a possible crack velocity. The arrows represent how a crack avalanches of slows down at the phase transition thresholds.
}
 \label{velmap}
\end{figure}
Equation (\ref{arrh3}) defines, for a given load $G$, a function $S_G$ such as: $V=S_G(V)$. To fit the model, the actual velocity at which a crack advances must be a solution of this equation (i.e., be a fixed point for the function $S_G$) \cite{dynamicalsystems}. Figure \ref{VSV} illustrates that, depending on the value of $G$, $S_G$ has one to three fixed points: three possible values for the crack velocity. This finite number of solutions arises from the steady-state approximation. If we were to consider the transient regimes, $S_G(V)$ would be, for a front propagating at any velocity $V$ and load $G$, a target velocity. Any crack not having reached a steady state would thus accelerate or slow down to follow this function. The intermediate fixed point, when it exists, is then unstable (virtually impossible): a crack with a velocity value just above this point ($V<S_G(V)$) is too slow to be steady. The heat generation at the tip is higher than what the diffusion can accommodate, the temperature rises and the velocity increases to converge to the upper fixed point. On the contrary, if a crack is slightly slower than the intermediate solution ($V>S_G(V)$), the crack cools down to the lower fixed point. We here assume that such transitions happen in a negligible time so the steady velocities are sufficient to describe the main dynamics.
The outer solutions of (\ref{arrh3}) being the only stable ones, the model displays a two-phase behavior. The lower velocity marks a slow phase. The temperature elevation at the crack tip has little effect on the propagation, as $\Delta T(V,G) \ll T_0$. The higher solution corresponds to a thermally weakened phase where $\Delta T(V,G)$ has reached the plateau temperature of Fig. \ref{ttovel}. The velocity is there increased as the induced heat is potentially significant compared to the thermal background.\\
Notice in Fig. \ref{VSV} that there are two particular values of the load $G$ for which either the lower or the higher phase ceases to exist. We denote them $G_\text{aval}$ and $G_\text{stop}$ (with $G_\text{aval}>G_\text{stop}$) as they correspond to mechanical loads at which a slow crack will have to avalanche to the thermally weakened phase or at which a fast (weakened) crack can only cool down to the slow phase. For $G$ in between these two thresholds, a hysteresis situation holds, there are several solutions for $V$ and the crack might or might not be thermally weakened, depending on the mechanical history. To $G_\text{aval}$ and $G_\text{stop}$ correspond some specific velocities $V_\text{aval}<V_\text{stop}$ in between which a crack cannot propagate, as any solution is there unstable. Figure \ref{velmap} shows the possible crack velocities for various values of $G$. One can notice how similar it is to a first order phase transition \cite{ThermodynamicsCP} for the order parameter $V$ associated to avalanches (jumps in $V$) triggered by variations in the driving field $G$ at temperature $T_0$. Such a description compares interestingly with various ($V$, $G$) branches that are experimentally reported, for instance in the rupture dynamics of pressure adhesives \cite{tape1, tape2}, PMMA \cite{PMA3ss, Lengline2011} or elastomers \cite{elastomers} and the model can hence be matched to actual data over decades of velocities \cite{TVD2}. Note that, in the hysteresis domain, we do not discriminate on the relative stability of each phase. One can however argue, by analogy with other phase transition systems \cite{ThermodynamicsCP}, that one of the two solutions could only be metastable, that is, in an equilibrium which is less energetically favorable than the one of the alternative phase. In this case, when traveling though an heterogeneous medium where the variations in fracture energy are enough to get shifts from only one state to the other, one of the phase could still be preferential for the crack propagation.
\begin{figure}
\includegraphics[width=1\linewidth]{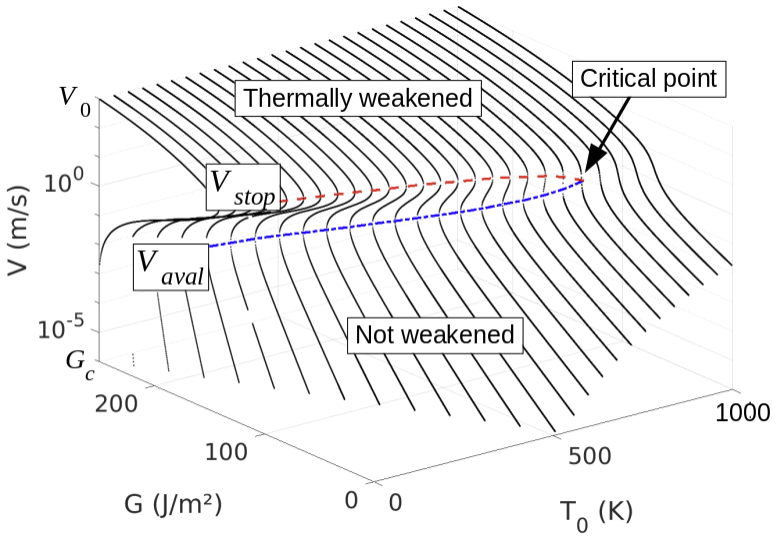}
  \caption{Solutions for the crack velocity as a function of $G$ and for various $T_0$. The dashed lines show the ($V_\text{stop}, G_\text{stop}$) and ($V_\text{aval}, G_\text{aval}$) couples and converge to the critical point.}
 \label{velmap2}
\end{figure}

\section{CRITICAL POINT}

Besides $G$, $T_0$ is the only other parameter of (\ref{arrh3}) which is not dependent on the medium's properties. Figure \ref{velmap2} thus shows the predicted propagation velocities for various ambient temperatures. Notice the existence of a critical ambient temperature: $T_0^*$, at which $G_\text{aval}=G_\text{stop}=G^*$ and $V=V^*$. Beyond $T_0^*$, the Joule effect cannot overcome the thermal background enough for the crack to be weakened. Increasing the load then only leads to a smooth increase in the velocity.
To relate to the theory of critical phenomena in phase transitions \cite{ThermodynamicsCP} we looked for the real numbers $\beta$, $\delta$ and $\gamma$ such that: 
\begin{equation}
\frac{V-V^*}{V^*} \sim \left(\frac{T_0-T_0^*}{T_0^*}\right)_{G=G^*}^{\beta} \\
\label{critexpo11}
\end{equation}
\begin{equation}
\frac{G-G^*}{G^*} \sim \left(\frac{V-V^*}{V^*}\right)_{T_0=T_0^*}^{\delta} \\
\label{critexpo12}
\end{equation}
\begin{equation}
\frac{G^*}{V^*}\frac{\partial V}{\partial G} \sim \left(\frac{T_0-T_0^*}{T_0^*}\right)_{G=G^*}^{-\gamma}
\label{critexpo13}
\end{equation}
where $\sim$ stands for a mathematical equivalence in the vicinity of the critical point (any pre-factor is overlooked). These exponents describe how $V$ converges towards $V^*$ beyond the critical point ($T_0 \ge T_0^*$). We also characterized how the hysteresis domain shrinks, looking for $\beta'$, $\delta'$ and $\gamma'$ such that:
\begin{equation}
\frac{V_\text{stop}-V_\text{aval}}{V^*} \sim \left(\frac{T_0^*-T_0}{T_0^*}\right)^{\beta'}
\label{critexpo21}
\end{equation}
\begin{equation}
\frac{G_\text{aval}-G_\text{stop}}{G^*} \sim \left(\frac{V_\text{stop}-V_\text{aval}}{V^*}\right)^{\delta'}
\label{critexpo22}
\end{equation}
\begin{equation}
\frac{G^*}{V^*}\frac{V_\text{stop}-V_\text{aval}}{G_\text{aval}-G_\text{stop}} \sim \left(\frac{T_0^*-T_0}{T_0^*}\right)^{-\gamma'}.
\label{critexpo23}
\end{equation}
With a bisection, we numerically estimated the critical point, checking for the number of solutions of $V=S_G(V)$ (three solutions below $T_0^*$ and one above). Analyzing the shape of the velocity map in the derived vicinity we found: $\beta \approx 1/3$, $\delta \approx 3$, $\gamma \approx 2/3$, and $\beta' \approx 1/2$, $\delta' \approx 3$, $\gamma' \approx 1$ (see the supplemental material). Both sets of exponents respect the scaling relation \cite{ThermodynamicsCP}: $2\beta+\gamma=\beta(\delta+1)$. We hence derived critical exponents which are, along the phase co-existence domain, the same as the mean field exponents for, say, the liquid-gas transition \cite{ThermodynamicsCP}, but different beyond the critical point. The mean field characteristic might arise from the statistical nature of the Arrhenius law only representing an average velocity while consecutive molecular bonds can be overcome at very different speeds. Another interpretation is that it translates the zero-dimensional character of our model. We have indeed disregarded any velocity variations and elastic interactions along the crack front, making the assumption that it is thin or symmetrical enough perpendicularly to the propagation direction.

\section{Simulations of 0D fronts\\in disordered media}

\begin{figure}[b]
  \includegraphics[width=1\linewidth]{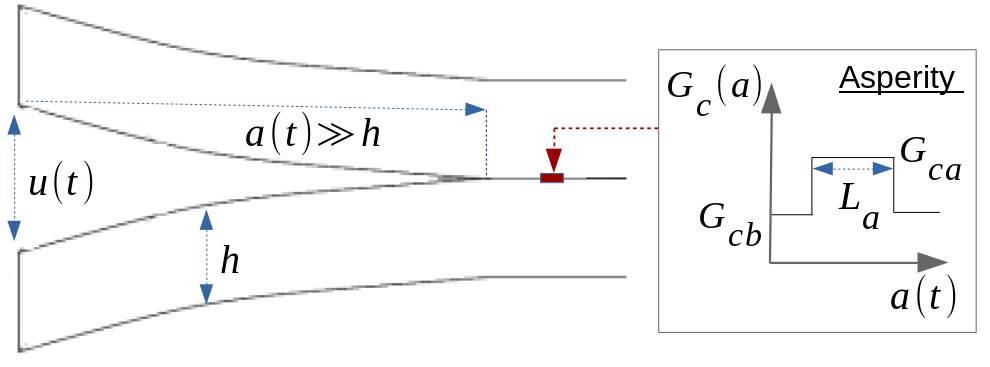}
  \caption{Geometry for the numerical simulations of zero-dimensional crack fronts overcoming a tough asperity.}
  \label{modeIload}
\end{figure}
\begin{figure}[b]
  \includegraphics[width=1\linewidth]{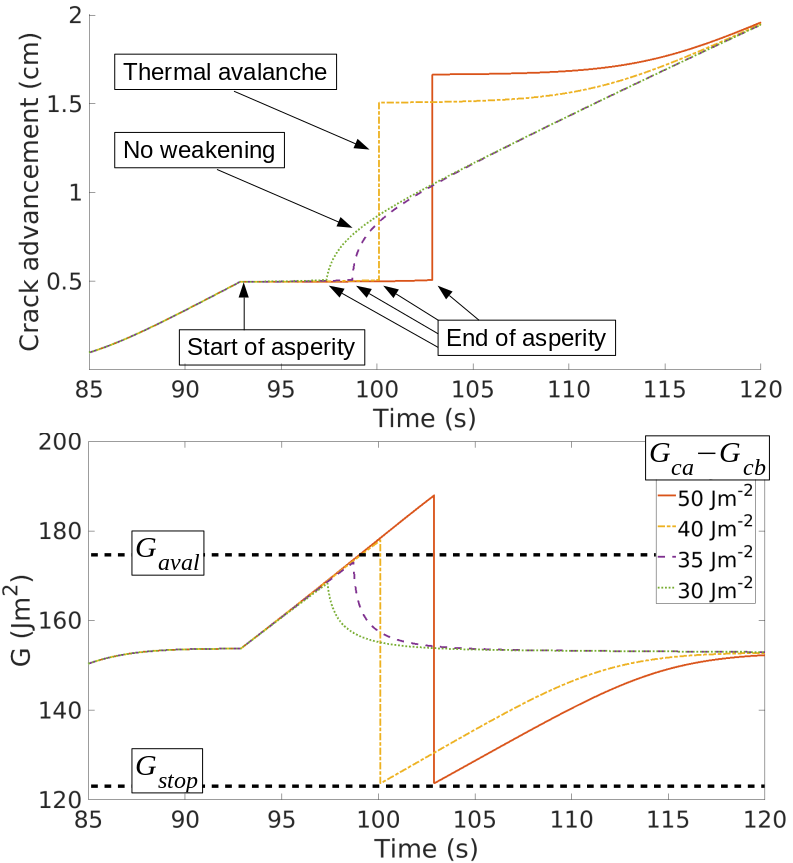}
  \caption{Numerical simulations for a crack overcoming an asperity as defined by the differential equation from (\ref{arrh3}) and (\ref{GmodeI}) and for various $G_{c_a}$. $L_a=100$\,$\mu$m, $v_u=120$\,$\mu$m s\textsuperscript{-1}, $h=5$\,mm and $E=3.2$\,GPa. The top plot is the crack advancement $a(t)$, the bottom one is the energy release rate $G(t)$. Thermal weakening is or is not triggered depending on the anomaly strength.}
  \label{obstacle}
\end{figure}
Let us finally illustrate the phase transitions with some simulations of such zero-dimensional fronts loaded in mode I. The loading geometry that we consider is shown in Fig. \ref{modeIload}. The support body consists in two sintered elastic plates which are progressively separated at the edge. The deflection on the side, $u(t)$ (in m), is increased linearly with time: $u(t)=v_ut$. Using the Euler–Bernoulli beam theory \cite{fracmec}, one can compute the energy release rate at the tip of such a system:
\begin{equation}
G(t)=\frac{3 Eh^3{v_u}^2t^2}{8{a(t)}^4} \hspace{0.5cm} \text{if } a \gg h,
\label{GmodeI}
\end{equation}
with $E$ the body Young modulus (in Pa), $h$ half of its thickness and $a$ the crack advancement such as: $V=\partial{a}/\partial{t}$. By inserting (\ref{GmodeI}) in (\ref{arrh3}), we obtain the differential equation in $a(t)$ that governs the crack progression and that we solved with a time step adaptive Runge-Kutta algorithm \cite{DormandPrince}.
We here consider a crack interface with a homogeneous background cohesion $G_c=G_{c_b}$ which is only disturbed by a single tough asperity of length $L_{a}$ ($G_{c_a} > G_{c_b}$). Figure \ref{modeIload} shows a schematic for this anomaly while Fig. \ref{obstacle} shows, for several values of $G_{c_a}$, the course of the crack over it and the corresponding evolution of the energy release rate. When the front reaches the asperity, the crack velocity dramatically decreases as it reaches a tougher area. Meanwhile the load $G$ increases because the far field deflection continues to build up on a now quasi-static crack. Once the anomaly finally gets passed, the simulations show two possible scenarios. If $G_\text{aval}(G_{c_b})$ (i.e., the phase shift threshold for the background $G_{c_b}$) was not reached over the anomaly, then the crack only accelerates back to its pre-asperity state. However, if $G_\text{aval}(G_{c_b})$ was overcome, the crack shifts phase and becomes thermally weakened: it avalanches until $G=G_\text{stop}$. In Fig. \ref{obstacle}, one can read the values of $G_\text{aval}$ and $G_\text{stop}$ and remark that they match the theoretical values displayed in Fig. \ref{velmap}. Note that, if the load was to be quickly increased, an avalanche could be triggered without the need for any asperity.  We showed, however, how the medium's disorder can lead to some spontaneous thermal weakening of the crack course.

\section{Discussion and conclusion}

By combining an Arrhenius law and the heat equation, we have thus demonstrated the possibility of a thermally activated dynamic phase transition in the propagation of cracks. This phase description may have major implications for the understanding of fracture dynamics. With a rather simple subcritical model, we indeed explain both slow creep regimes and fast ruptures. We do not however strictly disregard over-critical propagations, as $G>G_c$ only implies that the Arrhenius activation energy is null and hence always exceeded. In this case, we predict $V \sim V_0$. Note that at such high velocities, crack fronts tend to complexify \cite{RaviChandar1997, Fineberg_1991}, and our model might not hold as such, as it only considers single fronts. We derive a tip temperature approaching the $10^4$ K range. Although it is high, some experimental characterisations of triboluminescense \cite{tribo_blackbody, Bouchaud2012} have shown that fast cracks can reach such a temperature, which only stands on small volumes ($\sim l^2L$, where $L$ is the length of the front) and short time periods ($\sim l/V$) such that it does not imply a gigantic level of energy nor it necessary leads to local fusion or sublimation of the solid. Note that the temperature merely measures the amplitude of the atoms agitation, and that its statistical definition actually suffers for heat production zones smaller than the molecular scale. While atomic scale simulations \cite{Atom3} would be more appropriate to study the induced heat, such computationally demanding models are often run at given (fixed) temperatures. Yet, some occurrences \cite{shockfront_heat, AtomHeat} derive a non negligible induced heat.\\
Besides describing the two phases, we explained the potential shifts from one to the other and point out here how compatible this is with Maugis' reinterpretation \cite{Maugis1985} of the Griffith criteria \cite{Griffith1921} and so, with the usual stick-slip in brittle fracturing processes \cite{Santucci2004, tape1}, when avalanches get considerably larger than the scales of the in situ quenched disorder. We also showed that above a critical ambient temperature, $T_0^*$, this phenomenon cannot occur. For materials where $T_0^*$ is lower than the melting point at a given confining pressure, a same solid then displays a different behavior under cool or hot conditions: fragile when cold, but smoother/ductile when warm, as thermal avalanches are inhibited. The model thus could stand as a novel and physical explanation for the fragile-ductile transition of matter. Of course, it might be oversimplifying that to assume that all our parameters stay constant when varying $T_0$. The general physical principles however remain valid. Previous theories \cite{SiliconBDT, Atom2, BDTatom} actually support the importance of the crack-tip plasticity in the fragile-ductile transition, but rather relate it to the nucleation and mobility of dislocations ahead of the front. Such processes are compatible with induced thermal elevation \cite{shockfront_heat}, but are not directly captured by our mesoscopic description of the heat production zone.\\
Finally, and although we presented a mode I model, we suggest that some analogy is to be made with the frictional effects induced in mode II and mode III fracturing. Notably, as frictional heating is believed to be a cause for the instability of some seismic faults, a potential earthquake triggered when overcoming a strong fault plane asperity might indeed be amplified due to thermal weakening. The existence of the critical point would then explain the disappearance of such amplifications at higher depth (i.e., where rocks are in ductile conditions \cite{Scholz1988}) as the thermal background is there enough to make the frictional heating negligible and, hence, favors creep over brittle ruptures.\\

\section*{\label{sec:acknol} Acknowledgements\\and contributions}

\noindent T.V.-D. developed and analyzed the model and the simulations, and redacted the first versions of the manuscript. R.T. proposed the physical basis of the model and its mathematical formulation. A.C. set the basis for the numerical implementation of the model and the principles of the resolution algorithm. K.J.M. contributed in the interpretation of the model in fracture mechanics applications. E.G.F. contributed to analyze the model in terms of critical point characteristics. All authors participated to the redaction of the manuscript and agreed with the submitted version.
The authors declare no competing financial interests in the publishing of this work and acknowledge the support of the IRP France-Norway D-FFRACT, of the Universities
of Strasbourg and Oslo and of the CNRS INSU ALEAS program. Readers are welcome to comment and should address to vincentdospitalt@unistra.fr or renaud.toussaint@unistra.fr. See the Supplemental Material at [URL] for a discussion on the parameters we used, details on how to compute the temperature elevation at the crack tip and decades on which we fitted the critical exponents. 

\newpage
\FloatBarrier
\bibliographystyle{unsrtnat}
\bibliography{main.bib}

\end{document}


\title{Thermal weakening of cracks and brittle-ductile transition of matter: a phase model\\---------------------------------------------------------------------------------------------------------------\\SUPPLEMENTAL MATERIAL}

\author{Tom Vincent-Dospital}
\email{vincentdospitalt@unistra.fr\\}
\affiliation{Université de Strasbourg, CNRS, IPGS UMR 7516, F-67000 Strasbourg, France}
\affiliation{SFF Porelab, The Njord Centre, Department of physics, University of Oslo\\P. O. Box 1048, Blindern, N-0316 Oslo,  Norway}

\author{Renaud Toussaint}
\affiliation{Université de Strasbourg, CNRS, IPGS UMR 7516, F-67000 Strasbourg, France}
\affiliation{SFF Porelab, The Njord Centre, Department of physics, University of Oslo\\P. O. Box 1048, Blindern, N-0316 Oslo,  Norway}

\author{Alain Cochard}
\affiliation{Université de Strasbourg, CNRS, IPGS UMR 7516, F-67000 Strasbourg, France}

\author{Knut J\o rgen M\aa l\o y}
\affiliation{SFF Porelab, The Njord Centre, Department of physics, University of Oslo\\P. O. Box 1048, Blindern, N-0316 Oslo,  Norway}
\author{Eirik G. Flekk\o y}
\affiliation{SFF Porelab, The Njord Centre, Department of physics, University of Oslo\\P. O. Box 1048, Blindern, N-0316 Oslo,  Norway}

\date{\today} 
\keywords{Rupture dynamics, thermal weakening, statistical physics, critical phenomenon} 
                              
\begin{abstract}
In this supplemental material, we explain the parameters that we used for illustration in the main manuscript. We also provide analytical approximations for the temperature elevation at the crack tip and details how the critical exponents were derived. Although it is not essential to the core comprehension of our letter, we do refer to it. Therefore, for an easier understanding of the present material, we invite the reader to keep an eye onto the main manuscript, which references are marked with a `M'. For instance: Eq. (M1).
\end{abstract}     
                         
\maketitle
\tableofcontents
\newpage

\section{\label{sec:param}Chosen parameters (for illustration)}

Most of the parameters that we introduced in our model are strongly dependent on the medium in which the crack propagates. The chosen ones, for the figures of the main manuscript, use values we believe to be likely for the propagation of interfacial cracks in sintered polymethyl methacrylate (PMMA) bodies \cite{Lengline2011,Tallakstad2011}.
\citet{Lengline2011} have derived the main parameters of the Arrhenius law with experiments done at room temperature and for slow crack growth such that it is unlikely that a significant heat was released at the crack tip. In this experiments, $V$ was indeed reported to increase exponentially with $G$, meaning that $\Delta T(V,G)$ is negligible in Eq. (M3). They found $\alpha \sim 2.5 \times {10}^{-11}$\,m which is surprisingly less than the typical size of molecular bonds, one {\AA}ngstr{\"{o}}m. It was proposed \cite{Lengline2011}, as a possible explanation, that it is a consequences of the off-plane processes in the advance of a crack (a number of off-plane bonds have to be broken for the planar interface to advance by a projected length $\alpha$). Alternatively, it could be the translation that that only a part of the mechanical energy is consumed in breaking bonds \cite{Vanel_2009}. A nominal velocity $V_0 \times \exp[-\alpha^2 G_c/(k_B T_0)]$ was also measured, although the contributions of $V_0$ and $G_c$ were not individually resolved. We however used $V_0=1000$\,m\,s\textsuperscript{-1}, of the order of magnitude of the Rayleigh wave velocity \cite{Freund1972} ($\approx 1280$\,m\,s\textsuperscript{-1} in Plexiglas \cite{plexi}). This leads to $G_c=250$\,J\,m\textsuperscript{-2}. Note that to relate to the probabilistic molecular description of the Arrhenius law, one could also consider $V_0$ as the average molecular velocity from the kinetics theory \cite{kinetics}. If we approximate it as for unimolecular gas, $V_0 \sim \sqrt{8 k_B T_0/(\pi m)}$ where $m$ is the mass of a molecule), it ranges from $100$ to $1000$\,m\,s\textsuperscript{-1} for molecular masses from $100$\,g\,mol\textsuperscript{-1} (MMA molecule scale) to $10$\,g\,mol\textsuperscript{-1} (atomic scale).
Typical thermal constants for the Plexiglas were chosen as per the manufacturer specifications \cite{Plexiglas}: $C=1.7\times {10}^{6}$\,J\,K\textsuperscript{-1}\,m\textsuperscript{-3} and $\lambda=0.19$\,J\,s\textsuperscript{-1}\,m\textsuperscript{-1}\,K\textsuperscript{-1}. Finally, we chose a radius for the heat production zone of $l=20$\,nm corresponding to the scale of a few methyl methacrylate units, and we assume for simplicity that $\phi=1$: most of the energy contributes to the Joule effect and the other dissipation processes are comparatively negligible.\\
Note that such parameters are not accurately calibrated to match the actual rupture dynamics in interfacial PMMA. The purpose of this discussion is mainly to highlight the different orders of magnitude at stake in our presented model.

\section{\label{sec:computation} Temperature elevation at the crack tip}

\begin{figure}[H]
  \includegraphics[width=0.9\linewidth]{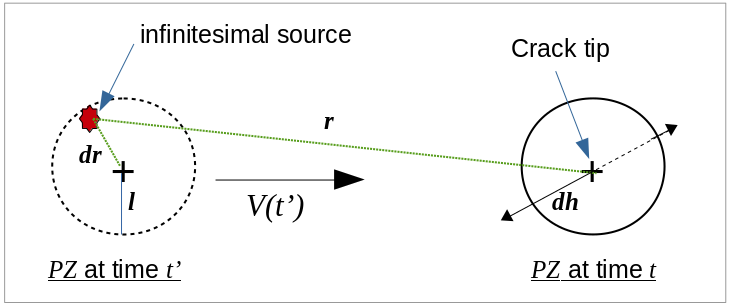}
  \caption{Illustration of the geometric parameters used in Eq. (\ref{green}). $dh$ is perpendicular to the figure perspective.}
  \label{croquis}
\end{figure}

As stated in the main manuscript, the temperature evolution at the moving crack tip is dictated by the following equation:
\begin{equation}
   \frac{\partial (\Delta T)}{\partial t} = \frac{\lambda}{C} \nabla^2 (\Delta T) + \frac{\phi G V}{C \pi l^2}f
   \label{tempdiff}
\end{equation}
which is a diffusion equation including a source term. $f$ is the support function of the thermal emission zone of surface integral $\pi l^2$ (we used $f=1$ in the zone and $f=0$ otherwise). It can notably be solved for the tip temperature using the diffusion Green's function \cite{HeatCond}:
\begin{flalign}
\begin{split}
&\Delta T(t)= \int_{-\infty}^{t} \mathrm{d}t' \iiint_{PZ(t')} \mathrm{d}v\, \frac{\phi G V}{C \pi l^2} \frac{\e^{\cfrac{-Cr^2}{4 \lambda (t-t')}}}{[\frac{4\pi\lambda}{C}(t-t')]^{\frac{3}{2}}} \\ \\
&\text{with } r=r(\mathbf{dr},t',t)=\| -\int_{t'}^{t} \mathbf{{V}} \mathrm{d}t'' + \mathbf{dr} \|
\label{green}
\end{split}
\end{flalign}
where $\mathrm{d}v$ is the volume of an elementary heat source of the advancing thermal process zone ($PZ$). This source is, at time $t'$, located at a distance $r$ from the crack tip at time $t$. We denote $\mathbf{dr}$ the positioning vector from the PZ center to the elementary source. See Fig. \ref{croquis} for an illustration of these parameters. The time integration of (\ref{green}) adds up the whole history of sources position, while the volumetric integral sums the contribution, for a given time $t'$, of infinitesimal heat sources of the emission zone.\\
The steady state which we used in our model is then the limit of $\Delta T(t)$ for large times ($t \longrightarrow +\infty$), assuming $V$ and $G$ are constant.

\newpage
\subsection{\label{sec:JouleHeating}Analytical approximation}

\begin{figure}[H]
  \includegraphics[width=1\linewidth]{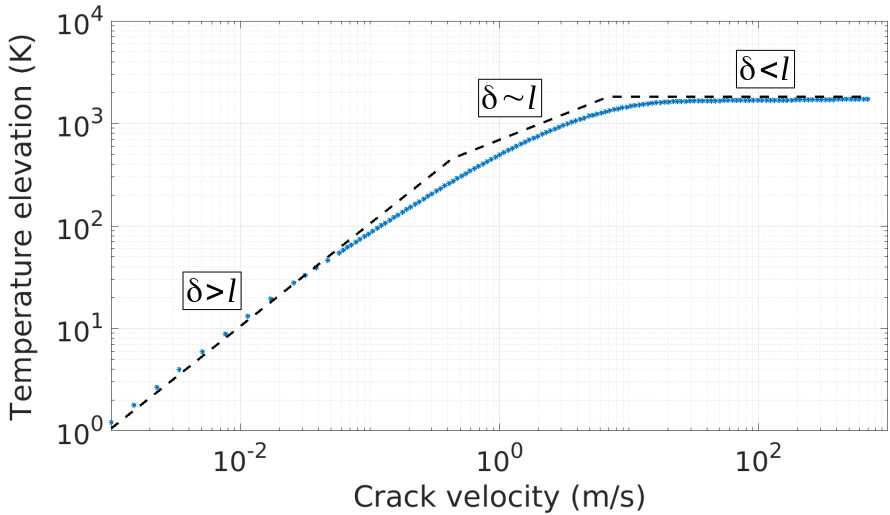}
  \caption{Steady state values of the temperature elevation (blue dots) obtained by solving the diffusion equation (\ref{tempdiff}) for a crack propagating at constant velocity for $\phi G = 200$\,J\,m\textsuperscript{-2}. The plotted asymptotes are the approximations (pre-factors included) from Eq. (\ref{velslow}) to Eq. (\ref{velmid}).}
  \label{ttovel2}
\end{figure}

For constant crack velocity and energy release rate, the temperature elevation at the steady state can be approximated \cite{ToussaintSoft}. When the crack propagates slowly, the temperature at the crack tip is constrained by the diffusion process. The heat production zone dissipates an energy $\phi G V \tau_0 dh$ during the time $\tau_0=l/V$ that it has spent over the tip location. $dh$ is the length of a crack front element. This energy is diffused over a roughly cylindrical volume with radius equal to the skin depth of diffusion $\delta$ over the time $\tau_0$. The section of such a cylinder is given by $\pi\delta^2 \approx \lambda\tau_0 /C$, leading to:
\begin{equation}
   {\Delta T}_{\text{slow}} = \frac{(\phi G V \tau_0 dh)}{C(\pi\delta^2 dh)} = \phi G\frac{V}{\lambda}.
   \label{velslow}
\end{equation}
In this case, because of the small velocity, $l$ is small compared to $\delta$. However, for cracks propagating fast enough $\delta \ll l$. The diffusion is then negligible and all of the energy stays in the emission zone of section $\pi l^2$:
\begin{equation}
   {\Delta T}_{\text{fast}} = \frac{(\phi G V \tau_0 dh)}{C(\pi l^2 dh)} = \frac{\phi G}{\pi C l}.
   \label{velfast}
\end{equation}
Note that ${\Delta T}_{\text{fast}}$ is the actual temperature at which the tip evolves in the weakened phase. Finally when $\delta \sim l$, the dissipated energy can only diffuse away from the heat production zone perpendicularly to the crack motion \cite{ToussaintSoft} and is spread over an ellipsoidal cross-section $\pi (\delta+l) l \approx 2\pi \delta l$, such as:
\begin{equation}
   {\Delta T}_{\text{mid}} = \frac{(\phi G V \tau_0 dh)}{C(2\pi \delta l dh)} = \phi G \sqrt{\frac{V}{4\pi C\lambda l}}.
   \label{velmid}
\end{equation}
Fig. \ref{ttovel2} illustrate the validity of these three approximations, showing the full solution of the diffusion equation (\ref{tempdiff}) together with asymptotes (\ref{velslow}) to (\ref{velmid}).
We show that, for slow cracks, the temperature increases linearly with the propagation velocity. At higher $V$, it however reaches a plateau constrained by the size of the heat production zone.

\newpage
\subsection{\label{sec:transienttime}Transient time}

\begin{figure}
  \includegraphics[width=1\linewidth]{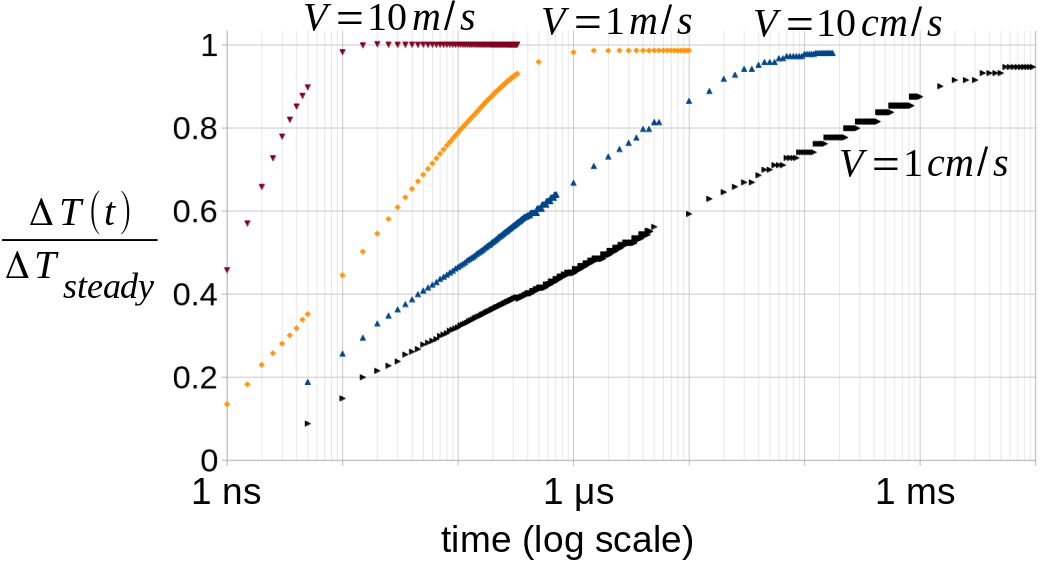}
  \caption{Transient response for the heating of a crack propagating at constant $V$ and $G$. The ratio of the tip temperature elevation and its steady state is shown for various velocities. One can read the actual value of $\Delta T_\text{steady}$ in Fig. \ref{ttovel2}.}
  \label{trans}
\end{figure}

Let us now move away from the thermal steady state hypothesis. We consider a tip which does not move before $t_0=0$\,s and propagates at velocity $V$ and energy release rate $G$ afterwards. Similarly to the steady state temperature, the transient time for its rise in temperature, $\tau$, is velocity dependent. For a fast propagating crack, the heat is not effectively evacuated away from the emission zone. The transient time then only corresponds to the maximum duration that the tip position stays in this zone:
\begin{equation}
\tau_\text{fast}(V) \sim \frac{l}{V}.
\end{equation}
For a slow crack, however, the characteristic time of the heat diffusion is to be considered. We can derive it by writing in Eq. (\ref{green}) that $r^2 \sim [V(t-t')]^2+l^2$ (see Fig. \ref{croquis}). When the first term of this sum dominates, the heat kernel behaves as $\exp[-(t-t')/\tau_\text{mid}]$ with:
\begin{equation}
\tau_\text{mid}(V) \sim \frac{4\lambda}{CV^2}.
\end{equation}
Overall, the quicker the crack progression, the shorter the thermal transient time. This is actually of convenience for our steady state approximation. For hot cracks, say propagating at velocities higher than $10$\,m\,s\textsuperscript{-1} (see Fig. \ref{ttovel2}), we have $\tau_\text{fast}<{10}^{-9}$\,s, which can be neglected for standard crack dynamics. Truly though, for slower cracks, say $V=1$\,cm\,s\textsuperscript{-1}, $\tau_\text{mid}$ gets in the millisecond range. In this case, however, we have $\Delta T \ll T_0$ and the Joule heating can anyway be neglected. For completeness, a third transient time also affects the thermal inertia. When the velocity is small enough for $r^2 \sim [V(t-t')]^2+l^2$ to be dominated by $l^2$, one can consider:
\begin{equation}
\tau_\text{slow} \sim \frac{Cl^2}{4\lambda}, 
\end{equation}
which in our case is a negligible ${10}^{-9}$\,s.\\
The transient regimes for various velocities are displayed in Fig. \ref{trans} for illustration. As mentioned, it corresponds to a cold crack that instantaneously accelerates from a full stop to a given constant velocity. It is of course a construction of the mind, as, in practice, our model predicts $V$ to evolve according to its Arrhenius subcritical growth. For instance, we explained in our manuscript how a velocity of $1$\,m\,s\textsuperscript{-1} corresponds to an unstable regime, so that what is shown in Fig. \ref{trans} for this velocity won't actually happen. The purpose this figure and of the expressions of $\tau_\text{fast}$, $\tau_\text{mid}$ and $\tau_\text{slow}$ is to give more insight on how quick the transition to stable regimes would be.
The discussion on how negligible are the transient regimes stays of course parameter dependent and one should keep in mind that our steady state model is an approximation that could hold more or less for different parameters values.

\section{\label{sec:MoreOnCP} More on the critical point}

\begin{figure}
  \includegraphics[width=1\linewidth]{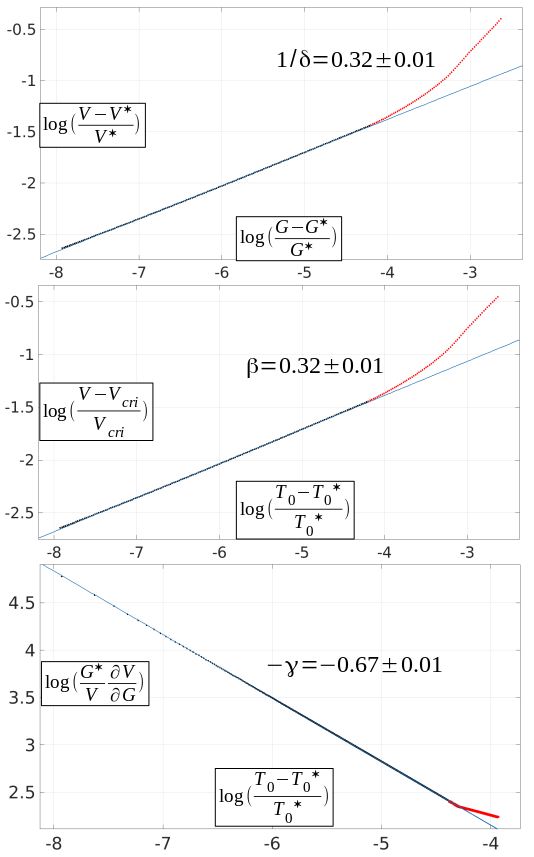}
  \caption{Representation in the log-log domains (base 10) of the $V$ surface (i.e., Fig. M4) along particular directions in the vicinity of the critical point. Black points were numerically derived from Eq. (M3) and the plain lines are the linear fits from which the red points were discarded. These fits are done beyond the critical point ($T_0>T_0^*$).}
  \label{decades}
\end{figure}
\begin{figure}
  \includegraphics[width=1\linewidth]{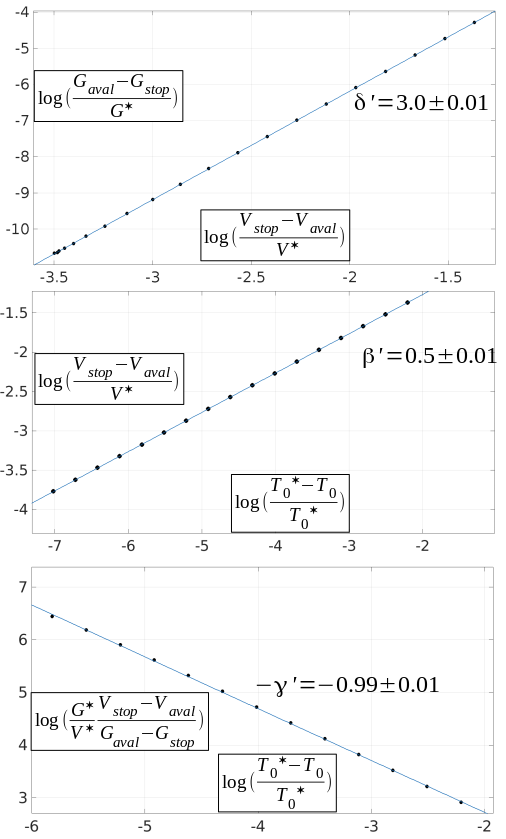}
  \caption{Representation in the log-log domains (base 10) of the $V$ surface (i.e., Fig. M4) along particular directions in the vicinity of the critical point. Black points were numerically derived from Eq. (M3) and the plain lines are the linear fits. These fits are done on the phase co-existence curve ($T_0<T_0^*$).}
  \label{decades2}
\end{figure}
We ran a bisection to numerically estimate the position of the critical point (CP). This bisection checked, for any temperature $T_0$, the maximum number of solutions of $V=S_G(V)$ to decide whether $T_0>T_0^*$ (a unique solution for any $G$) or $T_0<T_0^*$ (three solutions in the hysteresis domain). Numerical errors in this decision process imply some inaccuracy on the derived critical point. This, in return, leads to poorly determined critical exponents, as they are fitted in the direct neighborhood of the CP. We thus had to refine the position of the critical point with an iterative inversion in the vicinity of the firstly estimated location. The principle was to find $T_0^*$ and $G^*$ in this vicinity such that the velocity range on which to fit the exponents is maximized. The chosen procedure was to: first derive an a priori exponent $\beta_1$ in a direct neighborhood of the first CP estimation, then chose a smaller vicinity in which to vary the CP position, compute for each of these positions a more local $\beta_2$ value, derive a new CP position by minimizing $|\beta_2-\beta_1|$, iterate. Choosing smaller and smaller vicinities, this method allowed us to get a more accurate critical point and hence to expand the ranges on which to fit our final exponents. The corresponding decades and the respective exponents fits are shown in Fig. \ref{decades} and \ref{decades2}. We consider the reduced parameters $\Delta T_0/T_0^*$ and $\Delta G/G^*$, where $\Delta T_0=T_0-T_0^*$ and $\Delta G$ stands either for $G-G*$ or $G_\text{aval}-G_\text{stop}$ depending on the exponents (see Eq. (M4) to (M9)). Varying these parameters, the fits of the crack velocity function extend on at least three and a half decades (for $\beta$, $\delta$ and $\gamma$) and up to five decades (for $\beta'$ and $\delta'$).

\FloatBarrier
\bibliographystyle{unsrtnat}
\bibliography{scr.bib}